\documentclass[prx,twocolumn,showpacs,groupedaddress,superscriptaddress,nofootinbib,floatfix,preprintnumbers,longbibliography]{revtex4-2}
\usepackage{physics}
\usepackage{xcolor}
\usepackage{graphicx}% Include figure files
\usepackage{subfigure} % subfiguras
\usepackage{dcolumn}% Align table columns on decimal point
\usepackage{bm}% bold math
\begin{document}

\title{Principal frequency, super-bandwidth, and low-order harmonics generated by super-oscillatory pulses}
%\title{High harmonic generation driven by laser pulses with a super-bandwidth }% Force line breaks with \\
%\thanks{A footnote to the article title}%

\author{Enrique G. Neyra}
 %Lines break automatically or can be forced with \\
\email{enriquen@ciop.unlp.edu.ar}
   \affiliation{Centro de Investigaciones \'Opticas (CICBA-CONICET-UNLP), Cno.~Parque Centenario y 506, P.O. Box 3, 1897 Gonnet, Argentina}%\\This line break forced with \textbackslash\textbackslash

\author{Demian A. Biasetti}
\affiliation{Centro de Investigaciones \'Opticas (CICBA-CONICET-UNLP), Cno.~Parque Centenario y 506, P.O. Box 3, 1897 Gonnet, Argentina}%\\This line break forced with \textbackslash\textbackslash

\author{Fabi\'an Videla}
\affiliation{Centro de Investigaciones \'Opticas (CICBA-CONICET-UNLP), Cno.~Parque Centenario y 506, P.O. Box 3, 1897 Gonnet, Argentina}
\affiliation{Departamento de Ciencias B\'asicas, Facultad de Ingenier\'ia UNLP, 1 y 47 La Plata,Argentina}%\\This line break forced with \textbackslash\textbackslash

\author{Lorena Reb\'on}
\affiliation{ 
Departamento de F\'isica, FCE, Universidad Nacional de La Plata, C.C. 67, 1900 La Plata, Argentina%\\This line break forced with \textbackslash\textbackslash
}%
\affiliation{Instituto de F\'isica de La Plata, UNLP - CONICET, Argentina}

\author{Marcelo F. Ciappina}
\email{marcelo.ciappina@gtiit.edu.cn}
\affiliation{Physics Program, Guangdong Technion -- Israel Institute of Technology, Shantou, Guangdong 515063, China}
\affiliation{Technion -- Israel Institute of Technology, Haifa, 32000, Israel}
\affiliation{Guangdong Provincial Key Laboratory of Materials and Technologies for Energy Conversion, Guangdong Technion -- Israel Institute of Technology, Shantou, Guangdong 515063, China}
\date{\today}% It is always \today, today,
             %  but any date may be explicitly specified

\begin{abstract}

An alternative definition to the main frequency of an ultra-short laser pulse, named principal frequency ($\omega_P$), was recently introduced in E.G. Neyra, et al.~Phys.~Rev.~A \textbf{103}, 053124 (2021), resulting in a more transparent description of the nonlinear dynamics of a system driven by this coherent source. In this work, we extend the definition of $\omega_P$ incorporating the spectral phase of the pulse. This upgraded definition allow us to deal with super-oscillatory pulses as well as to characterize sub-cycle pulses with a complex spectral content. Simultaneously, we study the nonlinear interaction between a few-cycle super-oscillatory pulse with a gaseous system, analysing the spectral characteristics of the fundamental, third and fifth harmonics. Here, we make use of an \textit{ab-initio} quantum mechanical approach, supplemented with a wavelet analysis. We show that the spectral characteristics of the low-order harmonics are very well explained in terms of $\omega_P$, as well as the effective bandwidth of the super-oscillatory pulse. Our findings reinforce previous results that showed an increase of the effective bandwidth in the super-oscillatory region and the possibility to generate unique frequencies by a linear synthesis.  We open, thus, not only new perspectives in ultrafast optics, exploring novel pathways towards the generation of fully tunable strong and short coherent sources, but also discuss possible extensions of the concepts presented here to other wave phenomena, that can be found in acoustics, signal processing or quantum mechanics.

%, for instance, strong laser-matter interactions different kinds of spectroscopy, spectral and temporal, as well as in the strong-field ultrafast phenomena with the possibility to replacing the standard generation of the attoseconds pulses by these super-oscillatory pulses, which also, are possible to manipulate their effective bandwidth.

%the experimental realization of laser-matter nonlinear processes driven by super-oscillatory sources.

\end{abstract}

%\keywords{Suggested keywords}%Use showkeys class option if keyword
                              %display desired
\maketitle

\section{Introduction}

The
possibility to create and manipulate few-cycle ultra-short laser pulses, 
in the visible, infrared and XUV spectral ranges, has led to numerous studies in the field of laser-matter interaction, where this type of pulses finds diverse applications~\cite{brabec2000intense,liao2020towards,manzoni2015,liao2020towards,haessler2014optimization}. 
For instance, the interest on these pulses covers different topics, including the manipulation of the quantum properties of physical systems~\cite{krausz2014attosecond,rybka2016sub,morimoto2021asymmetric}, the study of the temporal dynamics of the chemical reactions~\cite{corkum2007attosecond, calegari2016advances} and the laser-matter interaction in the strong-field regime~\cite{amini2019symphony}.

Since the amplitude of the electric field varies within an optical cycle, the nonlinear matter interaction processes driven by few-cycle laser pulses, are strongly dependent on the carrier-envelope phase (CEP). This quantity plays an instrumental role in, for example, the high-order harmonic generation (HHG) phenomenon. This matter has been widely studied since the early days  (see e.g.~\cite{FerencReview2009} and references therein). 
Beyond this fact, we have recently showed in Ref.~\cite{NeyraPF2021} that,  in the few-cycle regime, the frequency that dominates the nonlinear matter interactions undergoes a small shift to frequencies higher than the carrier frequency of the pulse, $\omega_0$, usually considered as the main frequency to describe every nonlinear phenomena. This shift in the value of the main frequency is well represented by the so-called \textit{principal frequency}, $\omega_P$, a concept introduced in our previous work. According to its definition, $\omega_P$ corresponds to an increase with respect to the carrier frequency that depends on the bandwidth $\Delta \omega$ of the given pulse, i.e.~$\omega_P=\omega_P(\omega_0,\Delta \omega$).

On the other hand, in the last years there has been a great deal of interest in the study of the low-order harmonics generation (LHG). Here, the so-called near-threshold harmonics (NTHs) can be considered the most relevant ones. Their significance lies principally in the possibility to generate coherent light sources with high repetition rate and efficiency~\cite{chini2014coherent,yost2009vacuum}. While HHG is a well understood phenomenon, the theoretical description of LHG is rather more complicated and not exempt of controversies ~\cite{spott2014ab,spott2015transition}. Generally, it can be thought that the LHG can be explained by invoking the perturbation theory, meanwhile the usual HHG can be described by applying semi-classical arguments, e.g.~HHG is well understood employing the three-step model~\cite{Maciej1994}. When comparing with experimental measurements, however, one
finds that it is difficult to give a solid theoretical description
for the LHG. Firstly, perturbation theory is unsuitable when the driving laser is sufficiently intense. Secondly, due to the
important roles played by both the Coulomb potential and the
bound excited states, methods such as the strong field approximation (SFA), which (i) neglects the Coulomb potential in different parts of the description and (ii) only takes into account the ground state of the system, cannot be directly used to
describe the LHG. Thus, full quantum mechanical models, joint with those where the Coulomb potential is included, appears to be the most suitable ones to theoretically describe the underlying physics behind the LHG~\cite{Xiong2014,Xiong2017}.

In parallel, the phenomenon of super-oscillations, 
mathematically described by a band-limited function that can oscillate, locally, with a frequency higher than the highest frequency of its Fourier spectrum~\cite{aharonov2017,Berry_2006}, has raised a great deal of physical and mathematical interesting and intriguing implications~\cite{Berry_2019,zarkovsky2020transmission,chen2019superoscillation,rogers2020realising,colombo2022superoscillating,brehm2020temporal,ber2015superoscillations}. Indeed, in Ref.~\cite{Neyra_2021}, we have shown that the super-oscillatory phenomenon is also represented by laser pulses with a \textit{super-bandwidth}. When applied to a two-level system, and within the time window where the super-oscillation occurs, these pulses exhibit an effective bandwidth broader than the one fixed by the Fourier transform, which can open new ways in coherent control~\cite{he2019coherently,koong2021coherent}, or in ultra-fast spectroscopy.

In this work, we start by extending the definition of the principal frequency $\omega_P$, to take into account the spectral phase of the pulse, $\Phi(\omega)$. We show that this upgraded definition gives a more appropriate description of the main spectral contribution %, or central frequency, 
of super-oscillatory pulses. %,{Lorena: (esto no lo dir\'ia ac\'a) including one of the best known super-oscillatory function exhaustively studied in Ref.~\cite{Berry_2006,aharonov2017}}. 
In addition, we show that the same definition also allows a clearer description of the main frequency of sub-cycle pulses, with a very complex spectral content, like the one synthesized in Refs.~\cite{luu2015extreme,hassan2016}: the so-called \textit{optical attosecond pulses} or \textit{field transients}. Finally, we numerically study the bound-electron nonlinear response of a gaseous system driven by a super-oscillatory pulse, which can be implemented by means of an interferometric system. The study is carried out through the one-dimensional time-dependent Schr\"odinger equation (1D-TDSE) in an hydrogen atom. In particular, the spectral characteristics of the fundamental, third and fifth harmonics, in the temporal window where the pulse is super-oscillatory, were obtained through a wavelet analysis, from which we can conclude that the spectral features of these harmonics, i.e., their bandwidth and main frequency, are very well explained in terms of the super-bandwidth of the super-oscillatory pulses and the extended definition of the principal frequency.

\section{Principal frequency}\label{sec:intro}

We start by briefly discussing the alternative definition of the main frequency of an ultra-short laser pulse, that we have introduced in Ref.~\cite{NeyraPF2021}. Let us take the time-dependent electric field $E(t)$ of such a pulse, and its complex representation in the frequency domain $\tilde E(\omega)=\mathcal F[E(t)]=|\tilde E(\omega)|e^{i\Phi(\omega)}e^{i\phi}$, given by the Fourier transform $\mathcal F$, where $|\tilde E(\omega)|$ and $\Phi(\omega)$ are the spectral amplitude and phase, respectively, and $\phi$ is a global phase. %(the so-called carrier-enveloped phase (CEP)).
That frequency, which we have called the \textit{principal frequency} and denoted as $\omega_P$, is defined by the expression:
\begin{equation}
\omega_P=\frac{\int \omega^2 S(\omega) d\omega}{\int \omega S(\omega)d\omega},
\label{definition-Wp}
\end{equation}
where $S(\omega)=|\tilde E(\omega)|^2$ is the spectral power. Equation (\ref{definition-Wp}) can be seen as the mean of the spectral content of the laser pulse, weighted with a particular
density function, $\rho_P(\omega)=\omega S(\omega)$. In our previous work~\cite{NeyraPF2021}, we have shown that $\omega_P$ is related with the position of the maxima of $E(t)$, while the standard definition of the main frequency
$\omega_0=\frac{\int \omega S(\omega) d\omega}{\int  S(\omega)d\omega}$ \cite{BookUltrashort}, is related with its zeros.
This means that $\omega_P$ will offer a better description of the laser-matter interaction 
in the nonlinear regime, where the response of the system is led by some power of the peak field amplitude, $(E_0)^n$. 
However, a non-zero value of the difference $\omega_P-\omega_0$ can only be observed in the few-cycle pulses regime, where the envelope of the field changes significantly in an optical cycle. Furthermore, this difference becomes even larger for single-cycle and sub-cycle pulses (the latter are also called field transients).

It is evident that the definitions of both $\omega_0$ and $\omega_P$ only
take into account the spectral power $S(\omega)$ of the pulse and neglects its spectral phase $\Phi(\omega)$, which may be relevant in many cases. In particular, in the super-oscillatory (SO) phenomenon, which has an interferometric origin, the $\Phi(\omega)$ becomes instrumental. 
In this context, we introduce here an extended version of the principal frequency $\omega_P$, by considering in Eq. \eqref{definition-Wp} the field $\tilde E(\omega)$ itself instead of the spectral power $S(\omega)$:
\begin{equation}
\omega_{P_\Phi}\equiv\frac{\int \omega^2 |\tilde E(\omega)|e^{i\Phi(\omega)} d\omega}{\int \omega |\tilde E(\omega)|e^{i\Phi(\omega)}d\omega},
\label{e1}
\end{equation}
where the global phase $\phi$ was trivially canceled as it does not depend on $\omega$. Hence, we can understand the term $\rho_{P_\Phi}(\omega)=\omega |\tilde E(\omega)|e^{i\Phi(\omega)}$ in Eq.~\eqref{e1}, as a modified density function with respect to the one originally defined by Eq.~(\ref{definition-Wp}), which is now a complex-valued function. Consequently, $\omega_{P_\Phi}$ becomes also a complex quantity. 

In the following, we apply the new definition of the principal frequency $\omega_{P_\Phi}$ to some examples of SO functions as well as to describe the main frequency of an optical attosecond pulse.

\subsection{Super-oscillatory functions}\label{subsec:superoscillatory}

\begin{figure}[ht!]
\centering
\includegraphics[width=0.45\textwidth]{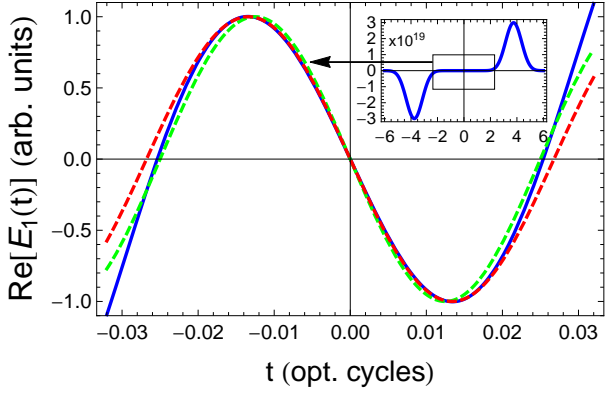}
\caption{Comparison between the SO pulse $E_1(t)$ (blue line), and sine waves of frequencies $\omega_{SO}$ (green dashed line) and $\omega_{P1_\Phi}$ (red dashed line), in the SO region (the central part of the pulse). In the inset plot we show the real part of $E_1(t)$ over several optical cycles ($2\pi/\omega_{f}$) of the sine waves, for the setting values $a=20$ and $N=15$.
} \label{fig:soexamplesB}
\end{figure}
As a first example, we analyse 
one of the best known SO functions, exhaustively studied in Refs.~\cite{Berry_2006,aharonov2017}, and experimentally synthesized in an ultra-short laser pulse in Ref.~\cite{Eliezer2017}. That SO field is described by the expression
\begin{equation}\label{e3}
E_1(t)=E_0\left[\cos(\frac{\omega_f t}{N})+i a \sin(\frac{\omega_f t}{N})\right]^Ne^{i\phi},
\end{equation}
where $N$ is a positive integer, $a$ is a real value greater than 1, and $\omega_f$ is the limit frequency of the Fourier spectrum, i.e., the frequency of the field when $\{N,a\}\rightarrow\{1,1\}$. For simplicity, in the following we normalize the amplitude of the field, setting its maximum value at $E_{0}=1$. The SO behavior is given by the fact that $E_1(t)\approx e^{ia\omega_f t}$, when $N>>1$ and for values of $\left |\omega_f t\right|/\sqrt{N}$ small enough, that is, in the central region, $E_1(t)$ oscillates with a frequency $\omega_{SO}=a\omega_f>\omega_f$.

In the frequency domain, the expression in Eq.~\eqref{e3} is a sum of Dirac delta functions:
\begin{equation}
\tilde E_1(\omega)=e^{i\phi}\sum_{j=0}^{N}C_j(N,a)\cross\delta\left(\omega-\omega_j(N)\right),
\label{e4}
\end{equation}
where
the discrete frequencies are given by $\omega_j(N)=\left(1-\frac{2j}{N}\right)\omega_f$, with Fourier coefficients $C_j(N,a)=\binom{N}{j}(\frac{1+a}{2})^{N-j}(\frac{1-a}{2})^{j}$. 
Then, the frequency $\omega_{P_\Phi}$ defined by Eq.~(\ref{e1}) %for the field $\hat{E_1}(\omega)$ (Eq.~(\ref{e4})), 
can be written as  
\begin{eqnarray}
\omega_{P1_\Phi}(N,a)&=&\frac{\sum_{j=0}^{N}\omega_j(N)^2C_j(N,a)}{\sum_{j=0}^{N}\omega_j(N) C_j(N,a)},
%\mathrm{and}\\
%\omega_{0_1}(a)&=&\frac{\sum_{j=0}^{N}\omega_j(N)C_j(N,a)}{\sum_{j=0}^{N} C_j(N,a)},\nonumber
\label{e5}
\end{eqnarray}
resulting in:
\begin{eqnarray}
\omega_{P1_\Phi}(N,a)&=&\left[\frac{1+a^2(N-1)}{aN}\right]\omega_f.
%\omega_{0_1}(a)&=&a~\omega_C.
\end{eqnarray}
%
%\begin{figure*}[ht!]
%\centering
%\includegraphics[width=0.4\textwidth]{figuraBerry.png}
%\caption{Comparison between the SO pulse $E_1(t)$ (blue line), and sine waves of frequencies $\omega_{SO}$ (green lines) and $\omega_{P1_\Phi}$ (red dashed lines). In left panel we plotted the real part of these pulses over several optical cycles of the sine waves, while in right panel we show a plot detail in the SO region. 
%} \label{fig:soexamples}
%\end{figure*}
%
This expression converges to 
$\omega_{SO}$, when $N \rightarrow \infty$.
Even more, for finite $N$ values, we can numerically demonstrate that $\omega_{P1_\Phi}(N,a)$ allows to describe the shape of the field, in the SO region, and localize its peak values (maxima and minima), in a better way than $\omega_{SO}$. 
This can be observed in Fig.~\ref{fig:soexamplesB}, where we show the real part of the field $E_1(t)$ for $\phi=\pi/2$, $a=20$ and $N=15$, and 
a zoomed image 
of the SO region where, 
for comparative purposes, we also plotted the harmonic waves oscillating at the SO frequency $\omega_{SO}$ ($\mathrm{Re}[e^{i a\omega_{f}t}e^{i\phi}]$), and at the frequency $\omega_{P1_\Phi}$  ($\mathrm{Re}[e^{i\omega_{P1_\Phi} t}e^{i\phi}]$).
It should be noted, however, that the position of the zeros of $E_1(t)$ are best described by $\omega_{SO}$. A more complete analysis for different values of the parameters $a$ and $N$, can be found in the Appendix~\ref{app}.

%
%From the last equations we can see that
%the frequency $\omega_{P_1}(N,a)$,  represents the ``frequency" of the maximum of the field, that is  $\omega_{P1}(N,a)=\omega_{0_1}(a)$ when $N \rightarrow \infty$. Additionally,
%the frequency $\omega_{0_1}(a)$ represents the usual approximation of the field $E_1(t)$ for the super-oscillatory frequency, that is, $E_1(t) \rightarrow e^{ia\omega_C t}$ when $N \rightarrow \infty$. 
%Additionally , the frequency $\omega_{P1}(N,a)$,  represents the "frequency" of the maxima of the field, that is  $\omega_{P1}(N,a)=\omega_{01}(a)$ when $N \rightarrow \infty$.

%As a second example, we study the SO functions proposed for modeling the combination of subdiffractive Gaussian-beams with an ultra-short Gaussian-pulse, which, as shown in Refs.~\cite{neyra2021tailoring, neyra2021simple}, can be obtained by means of simple interferometric techniques.

As a second example, we study the SO functions proposed to obtain subdiffractive Gaussian-beams and sub-Fourier ultra-short Gaussian-pulses, which, as shown in Refs.~\cite{neyra2021tailoring, neyra2021simple}, can be obtained by means of simple interferometric techniques. In the time domain, these SO pulses are the result of the interference between an ultra-short Gaussian pulse, characterized by a central frequency $\omega_C$ and a bandwidth  $\Delta\omega$, and the same pulse modified by a quadratic phase or a \textit{Gaussian filter}~\cite{neyra2021tailoring}. In the first case, a pulse $\tilde E_{G}(\omega)=e^{-(\frac{\omega-\omega_C}{\Delta\omega_2})^2}$ is the input of a Michelson interferometer, with a dispersive media in one of its arms, which introduce a quadratic phase (chirp) to the electric field traveling in that path. After being recombined into a beam splitter with the pulse that travels without modification, we can write the resulting SO field, in the frequency domain, as:
\begin{equation}
\tilde E_2(\omega)=e^{-(\frac{\omega-\omega_C}{\Delta\omega_2})^2}(1+\alpha_2 e^{i\theta(\alpha_2,\beta_2)}e^{i\beta(\omega-\omega_C)^2})e^{i\phi}.
\label{e7}
\end{equation}
Here, the value of $\alpha_2$ indicates the relative amplitude between the pulses in each arm of the interferometer, while $\beta_2$ is the quadratic chirp parameter. The condition of purely destructive interference, necessary to obtain SO pulses,
can be achieved by controlling $\alpha_2$ and $\beta_2$, and in consequence, the phase $\theta=\theta(\alpha_2,\beta_2)$~\cite{neyra2021tailoring}. 

For the pulse defined in Eq.~(\ref{e7}) the exact analytical expression of $\omega_{P_\Phi}$ is given by  
\begin{widetext}
\begin{eqnarray}\label{eq:e8}
\omega_{P2_\Phi}=\omega_C\frac{\left(1+\frac{\Delta \omega_2^2}{2\omega_C^2}\right)\Theta(\beta_2,\Delta\omega_2)+\alpha_2\;\mathrm{e}^{-i\tilde{\theta}(\alpha_2,\beta_2)}\left(1-\frac{i\Delta \omega_2^2}{2\omega_C^2\Theta(\beta_2,\Delta \omega_2)^2}\right)}{\Theta(\beta_2,\Delta\omega_2)+\alpha_2\;\mathrm{e}^{-i\tilde{\theta}(\alpha_2,\beta_2)}},
\end{eqnarray}
\end{widetext}

with $\tilde{\theta}(\alpha_2,\beta_2)=\theta(\alpha_2,\beta_2)+\frac{\pi}{4}$, and $\Theta(\beta_2,\Delta\omega_2)=\sqrt{\beta_2\Delta\omega_2^2-i}$. 
It is important to note that, when $\alpha_2 \rightarrow 0$, $\tilde{E}_2(\omega)$ approaches to the initial Gaussian pulse $\tilde{E}_{G}(\omega)$, and $\omega_{P2_{\Phi}}\rightarrow\omega_C\left[1+\frac{1}{2}\left(\frac{\Delta\omega_2}{\omega_C}\right)^2\right]$, which coincides with the expression that is obtained from the original definition of the principal frequency $\omega_P$ (Eq.~\eqref{definition-Wp}) introduced in Ref.~\cite{NeyraPF2021}, that does not consider the spectral phase of the field.

Alternatively to the previous interferometric system, 
a SO ultra-short pulse in the few-cycle regime, can be synthesized by introducing a filter with a rectangular spectral response in one of the arms of the interferometer, whose input is a $\mathrm{sinc}$-like pulse, instead of a Gaussian-pulse with a \textit{Gaussian filter} (see e.g. Ref.~\cite{neyra2021tailoring}).
In consequence, in the frequency domain, the resulting SO pulse is described by the expression 
 \begin{equation}
\tilde E_3(\omega)=\left[\mathrm{rect}\left(\frac{\omega-\omega_C}{\Delta\omega_3}\right)-\alpha_3~ \mathrm{rect}\left(\frac{\omega-\omega_C}{\beta_3 \Delta\omega_3}\right)\right]e^{i\phi},
\label{e10}
\end{equation}
where the initial pulse is represented by a rectangle function centered at the frequency $\omega_C$, with a bandwidth $\Delta\omega_3$, i.e., $\tilde E_{S}(\omega)=\;\mathrm{rect}\left(\frac{\omega-\omega_C}{\Delta\omega_3}\right)$. In Eq.~\eqref{e10},
$\alpha_3$ represents the relative amplitude between the pulses that travel through the different arms of the interferometer, just like in the previous interferometric scheme, while the parameter $\beta_3$ ($0<\beta_3<1$) is now the filtering parameter.  In this case, the purely destructive interference condition is easily reached introducing a $\pi$-phase between the arms of the interferometer. 
Therefore, by following the definition in Eq.~(\ref{e1}), we can see that $\omega_{P_{\Phi}}$ is now given by: %for $\omega_P(\omega_0,\Delta\omega,\alpha,\beta)$:
 \begin{equation}
\omega_{P3_{\Phi}}=\omega_C\left[1+\frac{1}{12}\left(\frac{\Delta\omega_3}{\omega_C}\right)^2 \left(\frac{\alpha_3\beta_3^3-1}{\alpha_3\beta_3-1}\right)\right].
\label{e12}
\end{equation}
Also in this case, when $\alpha_3 \rightarrow 0$ we recover the expression for $\omega_{P}$, i.e., 
$\omega_{P3_{\Phi}}\rightarrow\omega_{P3}=\omega_C\left[1+\frac{1}{12}\left(\frac{\Delta\omega_3}{\omega_C}\right)^2\right]$,when
$\tilde{E}_3(\omega)$ approaches $\tilde{E}_{S}(\omega)$~\cite{NeyraPF2021}.
\begin{figure*}[ht!]
\centering
\includegraphics[width=0.75\textwidth]{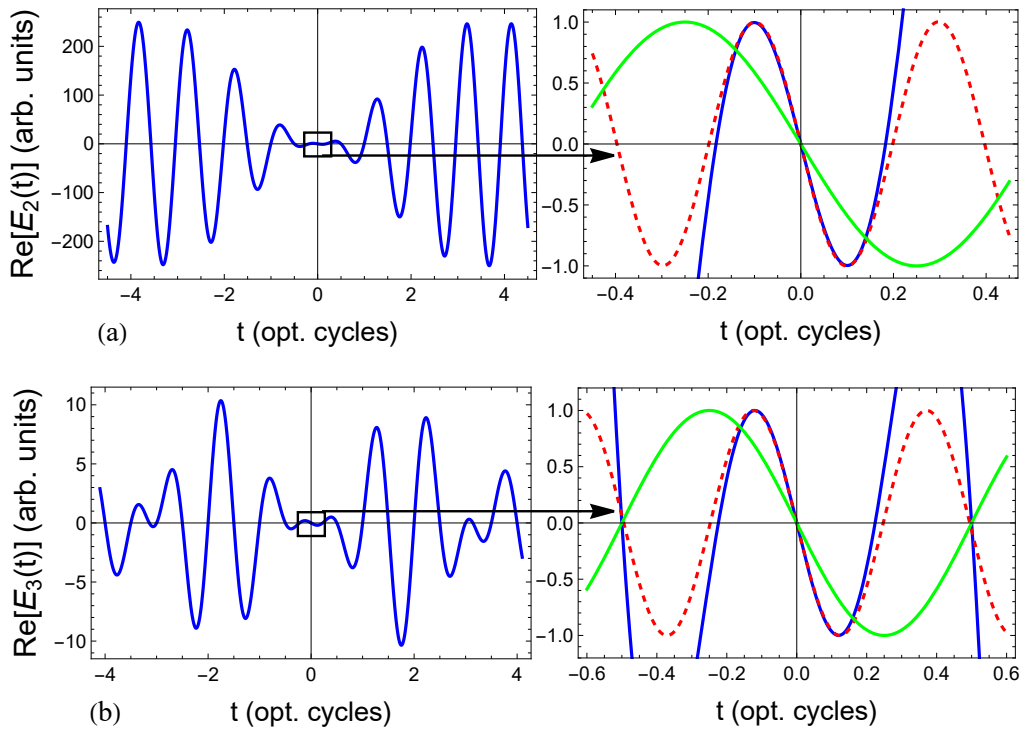}
\caption{Comparison between the SO pulses (blue solid lines) $E_2(t)$ (plot (a)) and $E_3(t)$ (plot (b)), and sine waves at the corresponding  frequencies $\omega_{Pi_\Phi}$ (red dashed lines) and $\omega_C$ (green solid lines), respectively. In the left panels, the real part of these fields ($\mathrm{Re}[E_i(t)]$, $\mathrm{Re}[e^{i\omega_{Pi_\Phi} t}e^{i\frac{\pi}{2}}]$, $\mathrm{Re}[e^{i\omega_C t}e^{i\frac{\pi}{2}}]$) are plotted over several optical cycles ($2\pi/\omega_{C}$) of the sine waves, while in the right panels a zoom over the SO region is depicted.} \label{fig:soexamples}
\end{figure*}

The SO behavior of both pulses, $\tilde{E}_2(\omega)$ and $\tilde{E}_3(\omega)$, becomes clear in the time domain. By Fourier transforming back Eq.~(\ref{e7}), we obtain the time-dependent description of $\tilde{E}_2(\omega)$:
\begin{eqnarray}
E_2(t)&=&\left(e^{-\frac{1}{4}(\Delta\omega_2 t)^2}+\frac{\alpha_2}{(1+\beta_2^2)^{1/4}} e^{-\frac{1}{4(1+\beta_2^2)}(\Delta\omega_2 t)^2}e^{i \Phi(t)}\right)\nonumber\\
&\times&e^{i \omega_C t}e^{i\phi}, \label{e8}
\end{eqnarray}
with the time-varying phase $\Phi(t)=\frac{\beta_2}{4(1+\beta_2^2)}t^2-\frac{1}{2}\arctan{(\beta_2)}+\theta$.
Analogously, from Eq.~(\ref{e12}), we obtain the time-dependent description of  
$\tilde{E}_3(\omega)$ as a sum of $\mathrm{sinc}$-functions with different temporal widths and amplitudes:
 \begin{equation}
 E_3(t)=\left[\mathrm{sinc}\left(\frac{\Delta\omega_3 t}{2}\right)-\alpha_3\beta_3~ \mathrm{sinc}\left(\frac{\beta_3\Delta\omega_3 t}{2}\right)\right]e^{i\omega_C t}e^{i\phi}.
\label{e11}
\end{equation}
Thus, $E_2(t)$ and $E_3(t)$ super-oscillate around $t=0$, %can be describe 
when the difference between the amplitudes of the interfering fields, i.e., the original pulse that enters the interferometer and the pulse that is broadening in one of its branches,
becomes small. This implies that $\frac{\alpha_2}{(1+\beta_2^2)^{1/4}}\rightarrow 1$, in the case of $E_2(t)$, and $ \alpha_3\beta_3\rightarrow 1$ in the case of $E_3(t)$. Besides, in this central temporal region, the amplitudes of $E_2(t)$ and $E_3(t)$ decrease as $\left(1-\frac{\alpha_2}{(1+\beta_2^2)^{1/4}}\right)$ and $(1-\alpha_3\beta_3)$, respectively.  For example, in the synthesis of $E_3(t)$, if the filtering parameter is chosen as $\beta_3=0.5$, the temporal FWHM of the pulse in the SO region approaches to zero when $\alpha_3 \rightarrow 2$, while $\omega_{P3_{\Phi}}\rightarrow \infty$. As expected, this will be accompanied by a significant reduction in the amplitude of the field: in fact, in that limit, the field amplitude decreases down to zero. The full theoretical description of this phenomenon was done in Ref.~\cite{neyra2021tailoring}.

In Fig.~\ref{fig:soexamples} we show the real part of $E_2(t)$ and $E_3(t)$ for the set of parameters:
$\phi=\pi/2$, $\Delta\omega_2=1$, $\alpha_2=2.24$, $\beta_2=5$, and $\Delta\omega_3=4$, $\alpha_3=1.95$, $\beta_3=0.5$, respectively. In these cases, we compare each SO pulse with harmonic waves oscillating at the corresponding frequency $\omega_{Pi_\Phi}$  ($\mathrm{Re}[e^{i\omega_{Pi_\Phi} t}e^{i\phi}]$) and at the central frequency $\omega_{C}$  ($\mathrm{Re}[e^{i\omega_{C} t}e^{i\phi}]$). In particular, because $\omega_{P2_\Phi}$ results in a complex value, we have plotted  $\mathrm{Re}[e^{i\mathrm{Re}[\omega_{P2_\Phi}] t}e^{i\phi}]$ instead of $\mathrm{Re}[e^{i\omega_{P2_\Phi} t}e^{i\phi}]$. Although we have not yet reached a final conclusion about the physical meaning of  $\mathrm{Im}[\omega_{P2_\Phi}]$, its value is very small in relation to $\mathrm{Re}[\omega_{P2_\Phi}]$ and its effect can be considered, to the first order, negligible. Thus, in the SO region a perfect description of $E_2(t)$ and $E_3(t)$ is obtained through monochromatic fields with frequencies $\mathrm{Re}[\omega_{P2_\Phi}]$ and $\omega_{P3_\Phi}$, respectively.

\begin{figure}[ht!]
\centering
\includegraphics[width=0.40\textwidth]{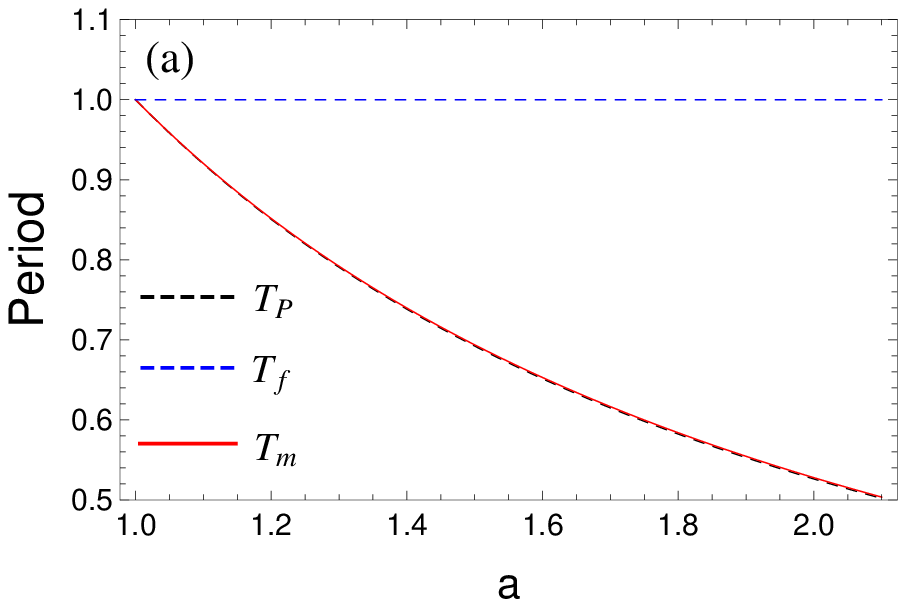}\\
\includegraphics[width=0.40\textwidth]{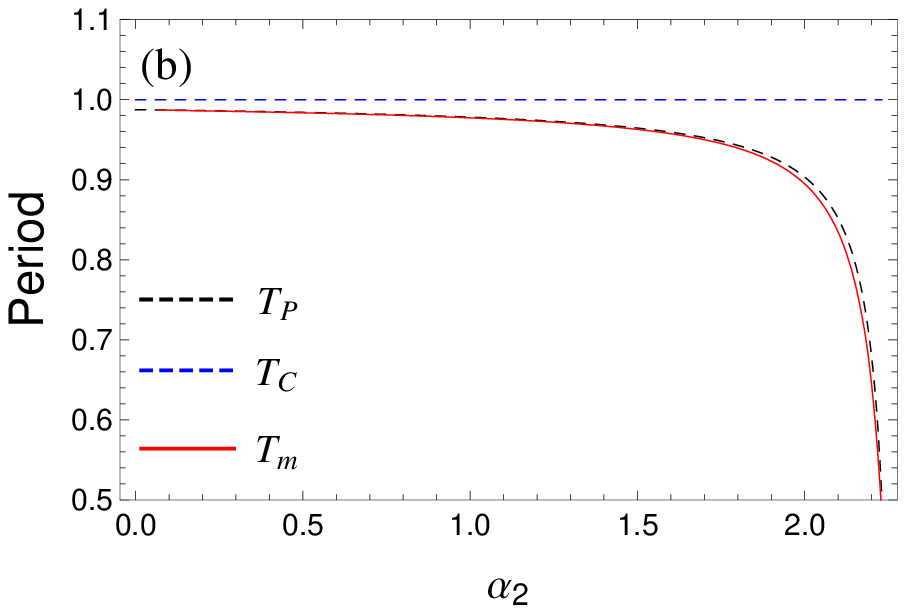}\\
\includegraphics[width=0.40\textwidth]{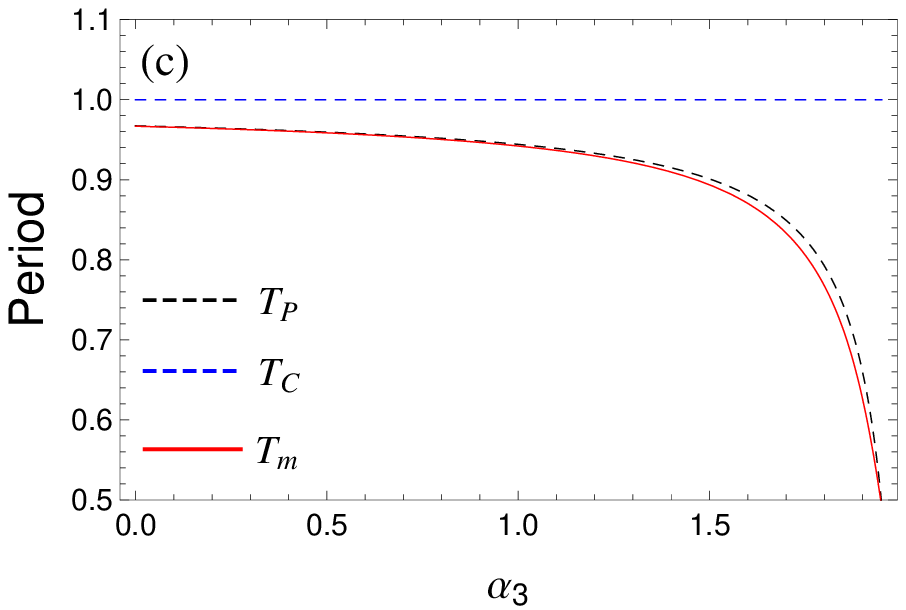}
\caption{Comparison between the principal period $T_P\equiv \mathrm{Re}\left[\frac{2\pi}{\omega_{P_\Phi}}\right]$ (dashed black lines) and twice the distance between a maximum and its adjacent minimum in the SO region, $T_{m}$ (red lines), as a function of the SO parameters for (a) $E_1(t)$, (b) $E_2(t)$ and (c) $E_3(t)$. The values are normalized with respect to the period $T_f=\frac{2\pi}{\omega_f}$ in the first case, and $T_C=\frac{2\pi}{\omega_C}$ in the other cases.} \label{fig:Tp-examples}
\end{figure}

For a deeper analysis of the use of $\omega_{P_{\Phi}}$ in the description of SO pulses, we numerically compare the \textit{principal period}, defined as $T_P=\mathrm{Re}\left[\frac{2\pi}{\omega_{P_{\Phi}}}\right]$, and twice the distance between a maximum and the adjacent minimum in the SO region, $T_{m}$. A similar analysis was made in Ref.~\cite{NeyraPF2021}, but considering the original definition of the principal frequency, $\omega_{P}$, and for a different type of ultra-short pulses in the few-cycle regime, which do not show a SO behavior.  
In Figs.~\ref{fig:Tp-examples}(a)--\ref{fig:Tp-examples}(c) we plot the values of $T_P$ (dashed black line) and $T_{m}$ (red line) for $E_1(t)$, $E_2(t)$ and $E_3(t)$, as a function of the SO parameters, $a$, $\alpha_2$ and $\alpha_3$, respectively. In the first case, the values of %SO parameters were change in such a way that the periods 
$T_{m}$ and $T_P$ were normalized to $T_f=\frac{2\pi}{\omega_f}$ (dashed blue line) with $\omega_f=2\pi$, while in the other two cases the normalization is done to the value $T_C=\frac{2\pi}{\omega_C}$ (dashed blue line), with $\omega_C=2\pi$. The excellent agreement between $T_P$ and $T_{m}$, indicates that $\omega_{P_{\Phi}}$ is an appropriate value for the frequency that dominates the SO region for the considered fields. 

\subsection{Principal frequency of a chirped pulse}
\label{app2}

In this sub-section we derive the principal frequency $\omega_{P\Phi}$ for a chirped Gaussian pulse with a quadratic phase, also known as group delay dispersion (GDD). 
In the frequency domain we can write this field as: 

\begin{eqnarray}
\tilde E_{\beta}(\omega)=e^{-(\frac{\omega-\omega_C}{\Delta\omega})^2}e^{i\beta(\omega-\omega_C)^2}.
\label{ebeta}
\end{eqnarray}
Equation~\eqref{ebeta} represents a chirped pulse characterized by a central frequency $\omega_C$, a bandwidth $\Delta\omega$, and  the chirp parameter $\beta$. 
Thus, from Eq.~\eqref{e1}, we arrive to the following expression:
\begin{eqnarray}
\omega_{P\beta{\Phi}}&=&\omega_C\left[1+\frac{1}{2}\left(\frac{\Delta\omega}{\omega_C}\right)^2 \left(\frac{1}{1+\Delta\omega^4\beta^2}\right)\right]\nonumber\\
&& -\;i\;\frac{\Delta\omega^4\beta}{2\omega_C(1+\Delta\omega^4\beta^2)}.
\label{ebeta2}
\end{eqnarray}
The pulse $\tilde E_{\beta}(\omega)$ can be seen as the pulse $\tilde E_2(\omega)$ of Eq. \eqref{e7}, when the control parameter $\alpha_2$ is $\gg 1$.
Just like in that case, $\omega_{P\beta{\Phi}}$ has a nonzero imaginary part and it is necessary to take its real part to characterize the main frequency of the pulse. Furthermore, when the chirp parameter $\beta \rightarrow 0$ the imaginary part vanishes and the expression in Eq.~\eqref{ebeta2} converge to the original definition of the principal frequency, $\omega_{P}$, for a Fourier limited Gaussian pulse, i.e., $\omega_{P\beta_{\Phi}}\rightarrow \omega_C\left[1+\frac{1}{2}\left(\frac{\Delta\omega}{\omega_C}\right)^2\right]$.

\begin{figure*}[ht!]
\centering
\includegraphics[width=0.40\textwidth]{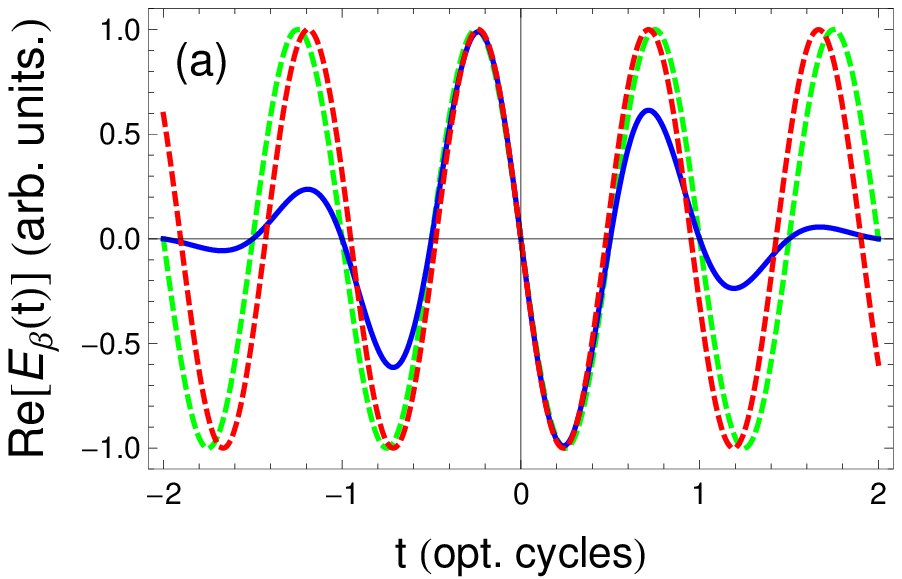}
\hspace{0.5cm}
\includegraphics[width=0.40\textwidth]{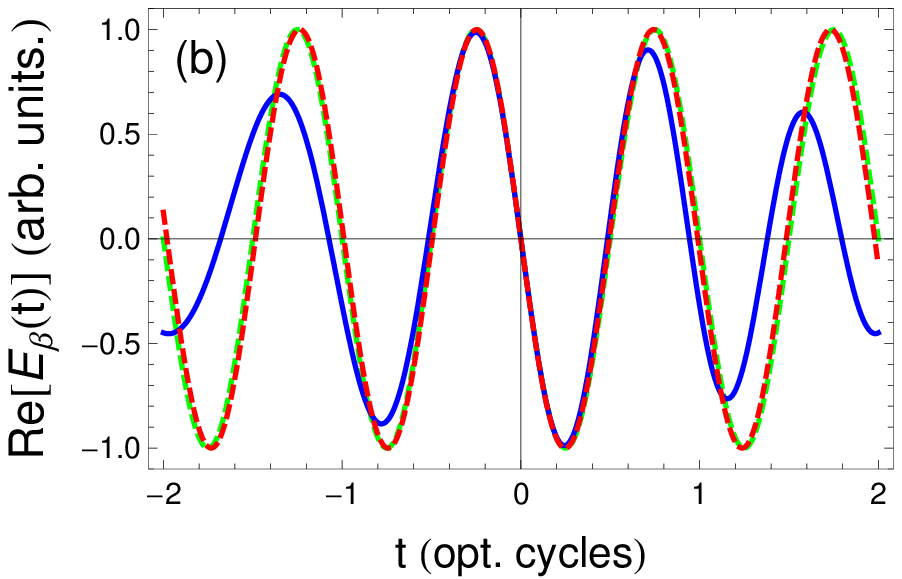}\\
\vspace{0.5cm}
\includegraphics[width=0.40\textwidth]{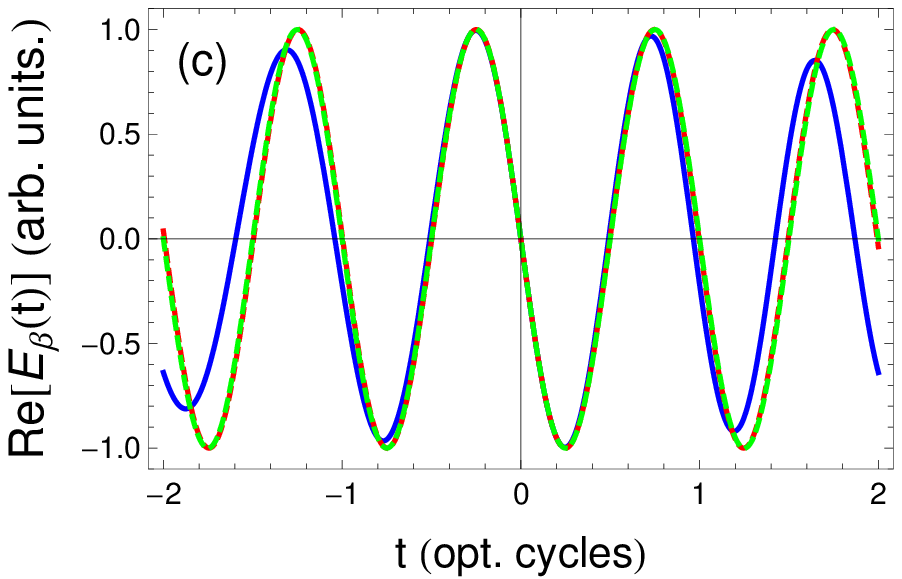}
\hspace{0.5cm}
\includegraphics[width=0.40\textwidth]{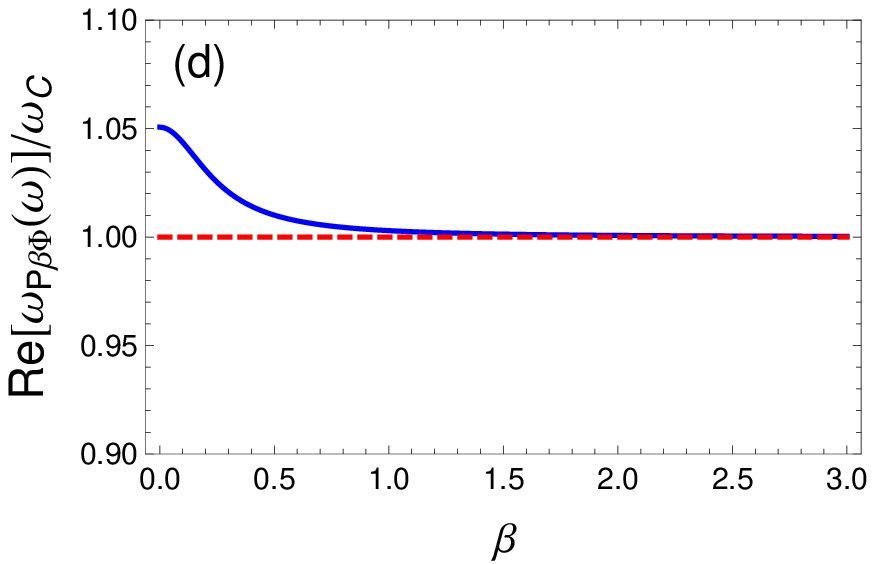}
\caption{(a) - (c) Comparison between a Gaussian chirped pulse $E_{\beta}(t)$ (blue line), and sine waves of frequencies $\mathrm{Re}[\omega_{P\beta{\Phi}}]$ (red dashed line) and $\omega_C$ (green dashed line). The values of the chirp parameter $\beta$ are 0 (a), 0.25 (b) and 1 (c). (d) Evolution of the ratio $\mathrm{Re}[\omega_{P\beta{\Phi}}]/\omega_C$ as a function of $\beta$ (blue line). The value of $\omega_C$ was normalized to the unity (red dashed line)} \label{wp-chirp}
\end{figure*}

In Fig.~\ref{wp-chirp} we show an example with $\omega_C=2\pi$ and  $\Delta\omega=2$, which corresponds to a temporal FWHM $\approx 1.18$ opt. cycles (FWHM $=4\frac{\sqrt{\log(2)/2}}{\Delta\omega}$). Figures~\ref{wp-chirp}(a), \ref{wp-chirp}(b) and \ref{wp-chirp}(c) correspond to the chirped Gaussian field (blue line) and sine monochromatic fields oscillating at frequency $\mathrm{Re}[\omega_{P\beta{\Phi}}]$ (red dashed line) and $\omega_C$ (green dashed line), respectively. It can be observed the temporal behavior of a Fourier-limited pulse ($\beta=0$) in Fig.~\ref{wp-chirp}(a), a chirped pulse with $\beta=0.5$ in Fig.~\ref{wp-chirp}(b), and a chirped pulse with $\beta=1$ in Fig.~\ref{wp-chirp}(c). Finally, in Fig.~\ref{wp-chirp}(d), we show the dependence of the principal frequency normalize to $\omega_C$, i.e. $\mathrm{Re}[\omega_{P\beta{\Phi}}]/\omega_C$, as a function of the chirp parameter $\beta$ (blue line). % where in red dashed line is represented the normalization of $\omega_C$. 
From this figure, it can be seen that $\mathrm{Re}[\omega_{P\beta{\Phi}}]$ decreases  monotonically with $\beta$, and converges to the central frequency $\omega_C$ when $\Delta\omega^4\beta^2\gg1$ (see Eq.~\eqref{ebeta2}). 
The last observation agreed with that is expected: for a big enough $\beta$ the chirped pulse becomes wider in the time domain and there is no longer an appreciable difference between its central frequency and the principal frequency. Thus, $\omega_{P\beta{\Phi}}\rightarrow \omega_{C}$ when $\omega_{C}>>\Delta\omega$.

\subsection{Optical attosecond pulse}

\begin{figure}[ht!]
\centering
\includegraphics[width=0.40\textwidth]{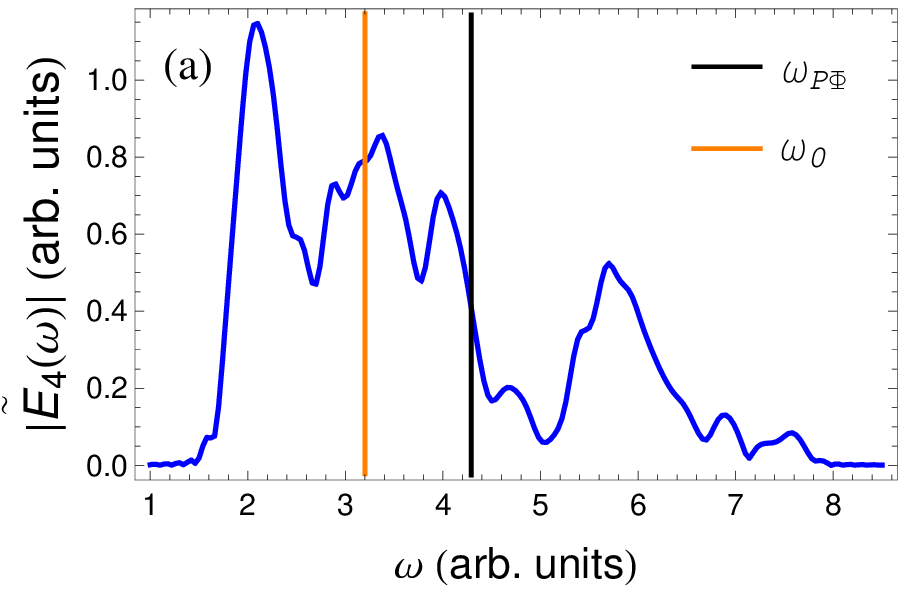}\\
\vspace{0.5cm}
\includegraphics[width=0.40\textwidth]{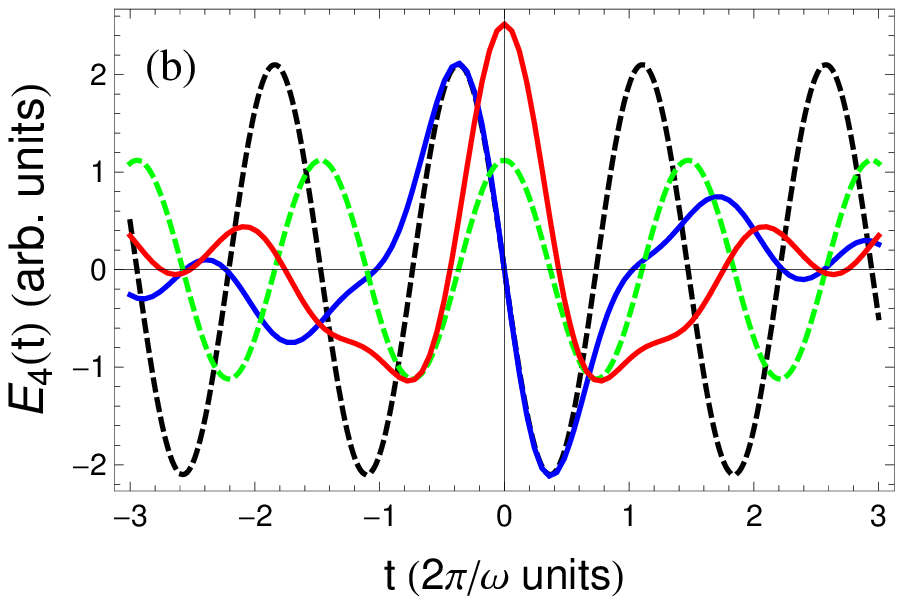}
\caption{(a) Spectral amplitude $|\tilde{E}_4(\omega)|$ of the optical attosecond pulse. (b) Temporal profile $E_4(t)$, with a cosine-like global phase (red line) and a sine-like global phase (blue line). Monochromatic fields oscillating at the frequency $\omega_{P4_{\Phi}}$ are plotted for a cosine-like (dashed green line) and a sine-like (dashed black line) phase.} \label{Exp-example}
\end{figure}

In this sub-section we use the definition of $\omega_{P_{\Phi}}$
to analyse an ultra-short pulse covering several octaves in frequency, which has been experimentally synthesized in Refs.~\cite{luu2015extreme,hassan2016}.
Here, the authors were able to synthesize a sub-cycle pulse in the visible spectrum, called \textit{optical attosecond pulse}, through a sophisticated experimental setup. A spectral amplitude $|\tilde{E}_4(\omega)|$, similar to that of such a pulse,  
can be seen in Fig.~\ref{Exp-example}(a). For this case, the spectral phase $\Phi(\omega)$ is constant so that   
we have numerically obtained the value of the principal frequency, directly from $|\tilde{E}_4(\omega)|$. This was done through a polynomial interpolation of the spectrum  $|\tilde{E}_4(\omega)|$ to subsequently calculate the integrals in Eq.~(\ref{e1}). In arbitrary units, we have obtained that $\omega_{P4_{\Phi}}=4.23$ (black line), while the value of its carrier frequency, obtained similarly from its definition \cite{BookUltrashort}, is $\omega_0=3.2$ in arbitrary units (orange line). 

The  optical attosecond pulse in the time domain $E_4(t)$ is shown in Fig.~\ref{Exp-example}(b), for both a sine-like global phase (blue line, $\phi=\pi/2$) and a cosine-like global phase (red line, $\phi=0$). 
We also show a monochromatic field oscillating at the frequency $\omega_{P4_{\Phi}}$, with a cosine-like global phase (dashed green line), and with a sine-like global phase (dashed black line). In the central region, it can be seen that $\omega_{P4_{\Phi}}$ allows to describe the position of the peak values of the field in both cases as well as the general shape of the pulse with $\phi=\pi/2$, before the first zero crossing. 

As mentioned in Section \ref{sec:intro}, one of the main motivations for extending the original definition of the principal frequency $\omega_P$ (Eq.~\eqref{definition-Wp}), was to include the
spectral phase $\Phi(\omega)$ of the field, which is relevant to described, for example, SO pulses.
For the attosecond optical pulse, however, since $\Phi(\omega)$ is constant, it does not play any role in the expression of $\omega_{P_{\Phi}}$ (Eq.~(\ref{e1})). Even so, $\omega_{P4_{\Phi}}$ differs from the value of the principal frequency $\omega_P$. In fact, in arbitrary units, we have obtained a value $\omega_{P}=3.63$, 
which represents a shift of $\approx$ 17$\%$,
towards lower frequencies with respect to $\omega_{P4_{\Phi}}$.
This difference arises from the fact that this extended definition of the principal frequency is weighted by the field $\tilde E(\omega)=|\tilde E(\omega)|e^{i\Phi(\omega)}$, instead of the spectral power $S(\omega)$, which allows to represent, in a much better way, the main frequency of ultra-short pulses with complex spectral content, with and without a particular spectral phase. Only when the spectral content of the pulse is symmetric (and $\Phi(\omega)$ is constant), as is the case for Gaussian or $\mathrm{sinc}$ pulses, the values of $\omega_{P}$ and $\omega_{P_{\Phi}}$ are equal.

\section{Principal frequency, super-bandwidth and LHG}\label{LHG}

In this section we analyse the nonlinear interaction between an atomic system and a SO pulse like the one given by Eq.~(\ref{e11}). For that purpose, after computing the LHG spectrum, we appeal to both the definition of $\omega_{P_\Phi}$, and the effective super-bandwidth of the pulse, %phenomenon, in laser pulses 
to characterize the generated harmonics. 
    
%In this section we compute the LHG spectra
%of an atom driven by SO laser pulses, 
%and appeal to both the definition of $\omega_{P_\Phi}$, and the effective super-bandwidth phenomenon, in laser pulses 
%to characterize it.
%In order to analyse the interaction between the bound-states of the atomic system and the SO pulse, we considered as driving field the one given by Eq.~(\ref{e11}) with fix values of the parameters $\beta_3$ and $\Delta\omega_3$, and a variable value of the parameter $\alpha_3$.

%In general, to compute the harmonics spectra in an atomic system it is necessary to find the quantum time dipole acceleration $a(t)$ and then, by the Fourier transform, is obtained the corresponding harmonics spectra, $S(\omega)=|\mathcal F[a(t)]|^2$. In our case, once obtained the time-dependent dipole acceleration, we study the spectral characteristic of the harmonics by a wavelet transform.    
As a first step, we calculate the dipole acceleration $a(t)$ quantum mechanically. $a(t)$ was obtained through the numerical integration of the one-dimensional time-dependent Schr\"odinger equation (1D-TDSE), in a hydrogen atom (for more details see e.g.~\cite{CiappinaPRA2012}).
After that, a time-frequency (wavelet) analysis has been performed to extract temporal information from the LHG spectrum. Here, we employ the Gabor transform to obtain $a_G(\Omega,t)$, defined as
\begin{equation}
a_G(\Omega,t)=\int dt' a(t')\frac{\mathrm{exp}[-(t-t')^2/2\sigma^2]}{\sigma\sqrt{2\pi}}\mathrm{exp}(i\Omega t'),
\end{equation}
where the integration is usually taken over the pulse duration, and $\sigma$ is chosen in such a way that a suitable balance between the time and frequency resolutions is achieved. In our case, we have used $\sigma=4/\omega_C$ for an appropriate resolution of the LHG spectrum.  

\subsection{Numerical results}
%
%%%%%% figure Gabor
\begin{figure*}[ht!]
\centering
\includegraphics[width=0.8\textwidth]{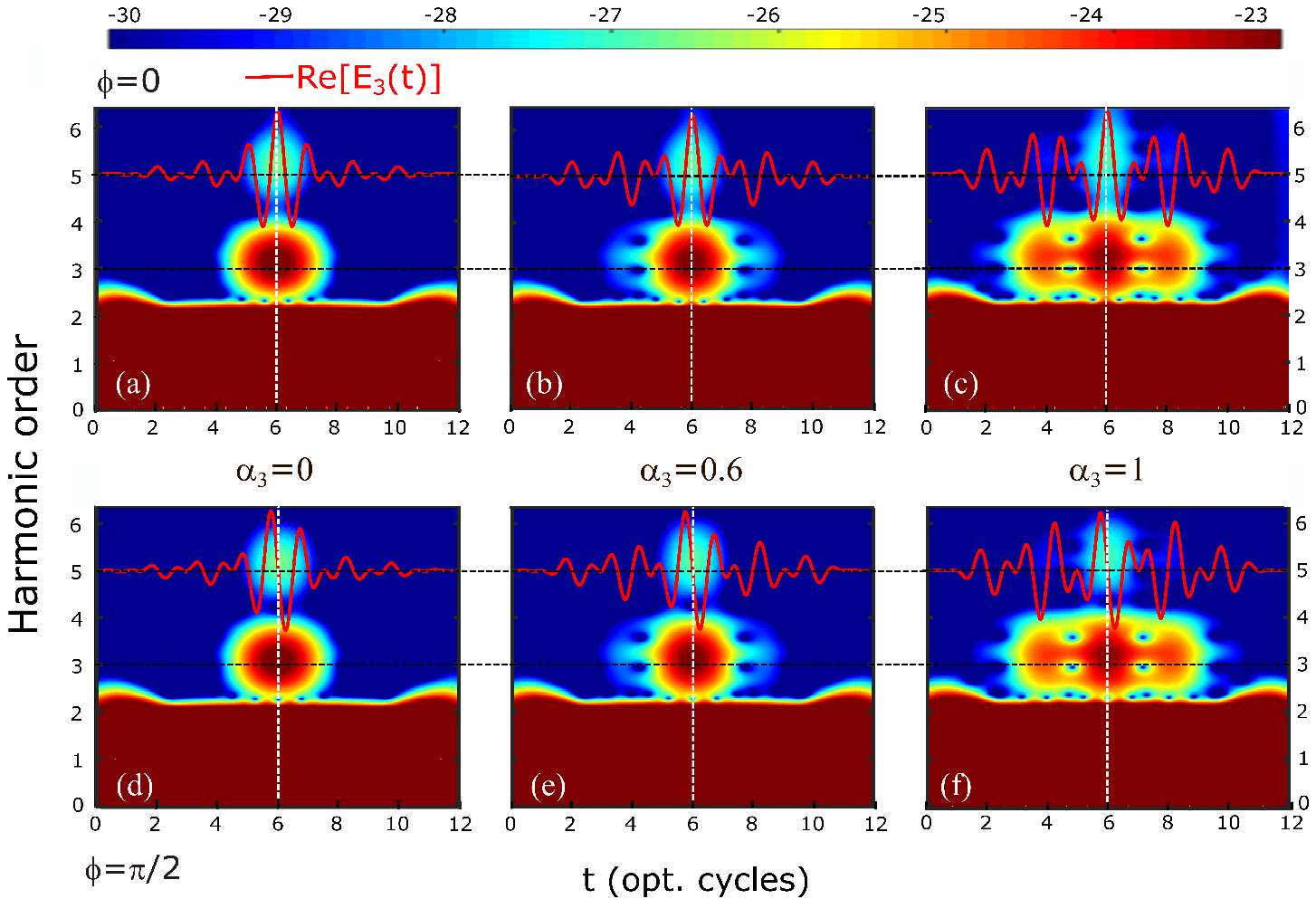}
\caption{Contour plot of $|a_G(\Omega,t)|^2$ (in log scale) as a function of time and the harmonic order, for different values of the synthesis parameter $\alpha_3$, related to the SO behavior of the driving field $E_3(t)$. From plots (a) to (c) we set the global phase of $E_3(t)$ to be $\phi =0$ while from plots (d) to (f) it is $\phi = \pi / 2 $. The values of $\alpha_3$ are indicated between the top and bottom panels. The corresponding synthesized pulse $E_3(t)$, used as driving field for the simulations, is included in each panel (red line), expressed in arbitrary units.}\label{fig:Gabor}
\end{figure*}
To perform our simulations we consider, as a driving field, the SO pulse $E_3(t)$ defined in Section II.A, with a central wavelength of $\lambda_C=1200$ nm, and a peak intensity $I_0=5\cross 10^{13}$ W/cm$^2$ in the SO region.  %We study six different fields $E_3(t)$, varying the parameter $\alpha_3$ at three values with their respective global phases, sine-like and cosine-like. 
The bandwidth, in units of $\omega_C=2\pi$, is set to $\Delta\omega_3=4$, and the values of the synthesis parameters are chosen to be $\beta_3 = 0.5$, while $\alpha_3$ is varied $(\alpha_3=0,0.6,1)$. It should be noted that for $\alpha_3=0$ we have a sinc-like pulse (Eq.~\ref{e11}), thus the temporal full width at half maximum (FWHM) is $\approx 5.564/\Delta\omega_3$ (see Ref.~\cite{NeyraPF2021}), which in the present example corresponds to 1.4 opt.~cycles. For $\alpha_3=0.6$ and $\alpha_3=1$, the value of FWHM in the SO region is reduced to 1.2 and 1 opt.~cycles, respectively. 
  
In Fig.~\ref{fig:Gabor} we show plots of $|a_G(\Omega,t)|^2$ (in log scale) for the different 
studied cases. The dependence on $t$ is expressed in optical cycles (opt.~cycles), while the dependence on the frequency $\Omega$ is derived from the order of the harmonic. The corresponding driving field $E_3(t)$
%synthesizes SO pulses (red line) are 
is overlapped to show the evolution of the radiation emission with the pulse duration. We observe that the third and fifth harmonics are emitted in the temporal region corresponding to the central region of the pulse (the SO region), %as two color circles who 
with an efficiency that decreases with the harmonic order. The fundamental harmonic, which lies in the saturated region in the lower part of the plots, has an intensity that is seven orders of magnitude greater than those of the third and fifth harmonics.
For $\alpha_3=0.6$ and $\alpha_3=1$, the radiation is emitted as harmonics in the temporal regions corresponding to the side-lobes of the pulse as well, which is typical of the SO phenomenon. %, but with less efficiency than in the central region. 
Indeed, the interference fringes can be understood as a consequence of the coherent superposition between the harmonic radiation generated by the SO region and that generated by the respective side-lobes. %It is worth mention that the fundamental harmonic is saturated in the contour plots, since its intensity is seven orders of magnitude greater than those of the third and fifth harmonics.

\begin{figure*}[!ht]
\centering
\includegraphics[width=\textwidth]{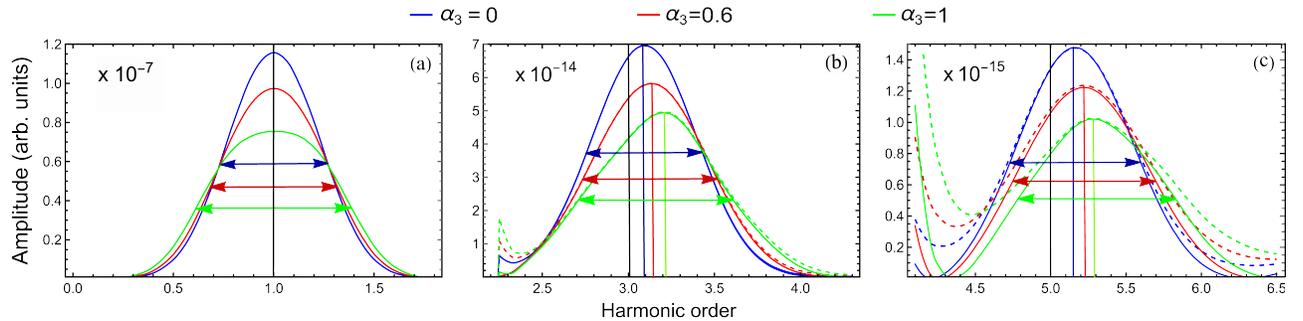}
\caption{Spectrum of the fundamental (a), third (b) and fifth (c) harmonic obtained from a vertical cut, in the SO region, of the contour plots of Fig.~\ref{fig:Gabor} (see there the white dashed lines). Each color indicates a different value of the synthesis parameter $\alpha_3$. Continue lines (dashed lines) correspond to a driving field $E_3(t)$ with a global phase $\phi=\frac{\pi}{2}$ ($\phi=0$).
Vertical color lines indicate the central frequency of the harmonics as quantified by the principal frequency $\omega_{P3_\Phi}(\alpha_3)$ as: $3\frac{\omega_{P3_\Phi}(\alpha_3)}{\omega_C}$ and $5\frac{\omega_{P3_\Phi}(\alpha_3)}{\omega_C}$, for the third and fifth harmonic, respectively. For comparison, black vertical lines indicate the central frequency of the third and fifth harmonics as quantified by $3\omega_C$ and $5\omega_C$, respectively, being $\omega_C$ the central frequency of the driving field. Horizontal arrows, indicate the temporal FWHM of the corresponding harmonic.} \label{harmonics}
\end{figure*}

For a better visualization of the evolution of the fundamental, third and fifth harmonics as a function of the synthesis parameter $\alpha_3$, we show in Fig.~\ref{harmonics}, a cross-section of the contour plots in the SO region (corresponding to the dashed white line in Fig.~\ref{fig:Gabor}). We present both the sine-like (continue lines) and cosine-like (dashed lines) phases. Two relevant phenomena can be extracted from this figure: 

\textit{(i)} There is a broadening of the harmonics bandwidth as $\alpha_3$ increases. For the fundamental harmonic, this result is equivalent to that one observed in Ref.~\cite{Neyra_2021}. This spectral broadening is quantified by the horizontal arrows in Fig.~\ref{harmonics}, indicating the FWHM of the harmonics. For the fundamental harmonic the FWHM is approximately 1.12 and 1.39 for $\alpha_3=0.6$ and  $\alpha_3=1$, respectively (here a FWHM equals to 1 corresponds to the FWHM of the driving field, i.e., when ~$\alpha_3=0$).
%where the widening for the fundamental harmonics is, approximately, 1.12 and 1.39 times, for $\alpha_3=0.6$ and $\alpha_3=1$ respectively, in relation to the initial pulse ($\alpha_3=0$). 
In the case of the third harmonic we have a broadening of 1.11 (for $\alpha_3=0.6$), and 1.2 (for $\alpha_3=1$), and for the fifth harmonic this broadening is 1.09 (for $\alpha_3=0.6$) and 1.19 (for $\alpha_3=1$).

\textit{(ii)} A blue-shift can be seen in the peak positions of the third and fifth harmonics, for all values of $\alpha_3$. Furthermore, this shift increases with $\alpha_3$.
For the fundamental harmonic, however, this phenomenon is not observed. To quantify this shift we compute the principal frequency $\omega_{P3_{\Phi}}$ for each set of values of the synthesis parameters $\{\Delta\omega_3=4,\beta_3=0.5,\alpha_3=0,0.6,1\}$, and calculated the third and fifth harmonics as $3\frac{\omega_{P3_{\Phi}}}{\omega_{C}}$ and $5\frac{\omega_{P3_{\Phi}}}{\omega_{C}}$, respectively. We have obtained that $3\frac{\omega_{P3_{\Phi}}}{\omega_{C}}\approx 3.10, 3.14,3.18$, 
while $5\frac{\omega_{P3_{\Phi}}}{\omega_{C}}\approx 5.17,5.22,5.30$, in order of increasing $\alpha_3$.
These values are represented by the vertical colored lines in Fig.~\ref{harmonics}, and it is clearly seen that they coincide with the central frequency of the respective harmonic order.

  Finally, it should be noted that the spectral characteristics of the LHG for different global phases $\phi$ (sine-like and cosine-like) of the driving pulse, shows that only noticeable changes are observed in the fifth harmonic. For $\alpha_3=0$ and $\alpha_3=0.6$, it can be seen a slight broadening in the spectrum for the cosine-like (dashed line) phase, in relation with the sine-like one, while for $\alpha_3=1$ the spectrum broadens towards a continuum (dashed green line). These results are in agreement with the nonlinear response of bound electrons studied with sub-cycle pulses Ref.~\cite{hassan2016}.  

\section{Discussion}

Here we discuss the physical implications and the consequences of the results presented in Section \ref{LHG}. We also put forward some questions that naturally remain open, since they need a deeper analysis beyond the possible practical applications and the scope of the present work.    

Firstly, 
the introduction of the principal frequency $\omega_P$ turns out to be instrumental to identify the frequency that dominates the interaction between an ultra-short pulse and matter, in the nonlinear regime, as well as to be 
the key parameter in describing the main frequency of SO pulses. 
Previously, in Ref.~\cite{NeyraPF2021}, we studied the HHG in atoms. In that work, we showed that the more energetic photon that is generated (cutoff) is better predicted by $\omega_P$, instead $\omega_0$. Besides, the extended version of $\omega_P$ introduced in the present work, and which we refer as $\omega_{P_\Phi}$, allows, on the one side, a very good description of the main frequency of ultra-short pulses with complex spectral content like the \textit{optical attosecond pulse}. On the other side, since this extended definition incorporates the spectral phase $\Phi(\omega)$, a good description of the SO frequency is obtained. In this regard, we have shown that a monochromatic field with frequency $\omega_{P\Phi}$ satisfactorily describes the respective time-dependent SO field. To the best of our knowledge, this is the first time that a mathematical expression to quantify the SO frequency has been presented.  

In relation to the blue-shift of the central peak of the LHG %harmonics considered 
observed in the present work, we can introduce the following analysis: in perturbation theory, the nonlinear polarization $P_{NL}(t)$, for an instantaneous response and a centrosymmetric medium, relates to the driving field $E(t)=A(t)e^{i\omega_C t}$ as $P_{NL}(t)$= $\chi_3E(t)^3$+$\chi_5E(t)^5$, where the parameters $\chi_3$ and $\chi_5$ are the electrical nonlinear susceptibilities (we consider here up to the fifth order). This implies that the harmonic fields $E_3(t)$ and $E_5(t)$ can be written as: $E_3(t)=A_3(t)e^{i\omega_{C3} t}$ and $E_5(t)=A_5(t)e^{i\omega_{C5}t}$, with $A_3(t)$ and $A_5(t)$ the fields envelopes, and central frequency $\omega_{C3}=3\omega_C$ and $\omega_{C5}=5\omega_C$, respectively. However, our results show that the central frequencies of the harmonics are $\omega_{C3}=3\omega_{P_\Phi}(\Delta\omega,\omega_C)$ and $\omega_{C5}=5\omega_{P_\Phi}(\Delta\omega,\omega_C)$, i.e, they are not only  dependent on the central frequency of the driving field $\omega_C$, but also on its bandwidth $\Delta\omega$. These results are equivalent to those presented in Ref.~\cite{shcherbakov2019photon}, where the blue-shift is originated by a photon acceleration phenomenon, which could be related with the properties of the nonlinear medium. In our case, on the contrary, the blue-shift is given by the principal frequency that is a characteristic of the laser pulse itself and is, therefore, independent of the medium. This fact would allow to apply the principal frequency concept in other nonlinear processes, for instance, the above-threshold ionization (ATI) in atoms~\cite{Milosevic2006} or HHG in solids~\cite{HHGSolid2022}, just to name a few.

The original idea behind the introduction of the principal frequency was to give a higher ``weight" to the more energetic frequencies (or photons) in the density distribution $\rho_P(\omega)=\omega S(\omega)$. The extended version of the density distribution $\rho_{P_\Phi}(\omega)=\omega |\tilde E(\omega)|e^{i\Phi(\omega)}$, is a complex-valued function that also gives a higher weight to the more energetic photons, but now it takes into account the spectral phase of the field. A question that arises is, what is the real physical meaning of the complex-valued principal frequency? In the temporal domain, the distribution  $\rho_{P_\Phi}(\omega)$ is the derivative of the field $\frac{dE(t)}{dt}$, that can be obtained using Fourier transform properties. Therefore, is there a relation between $P_{NL}(t)$ and $\frac{dE(t)}{dt}$? 

%In addition, the principal frequency works well for non-lineal phenomenon, which are not represented by the Fourier transform. That mean, that the principal frequency is outside the Fourier formalism?    

Second, beyond the concept of the principal frequency, the most relevant result presented in this work is the possibility to generate \textit{new frequencies or photons} in a laser pulse through a linear synthesis, lacking the need of a material medium. Although the concept of \textit{super-bandwidth} in laser pulses was presented in Ref.~\cite{Neyra_2021}, through the analysis of the stationary points in a two-level system, the study done in the present work is much more powerful and universal. This is because we have used \textit{ab initio} tools, the solution of the 1D-TDSE, and the analysis of the emitted radiation by a wavelet transform. Moreover, the \textit{super-bandwidth} appears here as a broadening of the bandwidth of the studied harmonics and is in excellent agreement with the value extracted from the dependence of the principal frequency $\omega_{P_\Phi}$ with the bandwidth.

One of the questions that can be gleaned from the previous paragraph is: Which is the limit  for the \textit{new frequencies or photons} that can be generated by this interferometric method? A similar question, \textit{How can an infrared photon behave as a gamma ray?}, was analysed in Ref.~\cite{berry2017escaping}. Theoretically, it seems that there is not such a limit, but experimentally it is necessary an extremely precise control of the phase between the interferometer arms, in order to maintain the condition of destructive interference. In addition, for a fixed value of the filtering parameter $\beta_3$, it is necessary, as well, a precise control of the relative field amplitudes between the interferometer arms, $\alpha_3$.         

Finally, what happens with the global phase $\phi$, in the SO region, when $\omega_P\gg\omega_C$ (i.e., the temporal FWHM $\rightarrow 0$)? In the SO region, are there changes in the spectral properties of the pulse when $\phi$ varies in that limit? 

\section{Conclusions}

In this work, we have extended the definition of the principal frequency by changing the \textit{weight} function in the spectral power $S(\omega)$ 
of a field $\tilde{E}(\omega)$, including now its spectral phase $\Phi(\omega)$. We show that, for the SO pulses presented as examples, this redefinition gives a frequency value that allows a well description of the main frequency in the SO region. %, one of them is the most know SO function. 
In addition, we have demonstrated that the principal frequency describes the main frequency of an optical attosecond pulse, with a very complex spectral content. We can conclude that the concept of principal frequency can be more suitable in the characterization of the main frequency of pulses emerging from complex synthesis techniques, instead of the conventional central frequency (wavelength), which is the standard experimental parameter used for that purpose (see for example Refs.~\cite{luu2015extreme,hassan2016, ridente2022electro,alqattan2022attosecond,huang2011high}).

Simultaneously, we have analyzed the nonlinear interaction properties of SO pulses, in the few-cycle regime, with a hydrogen atomic system, by solving the 1D-TDSE. Particularly, we studied the fundamental, third and fifth harmonics trough a wavelet analysis, and showed that their spectral characteristics are well described by resorting to the principal frequency and the effective super-bandwidth of the pump pulse. Finally, we made a discussion of the physical implications of our results, while some questions remain open.

The possibility to manipulate the spectral and temporal characteristics of a pulse in the SO region could be of great interest for a wide variety of applications, ranging from coherent control to femto/attosecond spectroscopy. On the other hand, the broadening of the fundamental harmonic as a \textit{super-bandwidth}, clearly shows that pulses in the SO region interact, in the weak field regime, as if they would have new frequencies or photon energies, i.e, new frequencies are being generated through a linear synthesis. In the same way that the super-oscillatory phenomenon is a wave phenomenon, this analysis can be extended to all kind of signals as, e.g.~acoustic waves, coherent sources, matter waves, etc.

The case of a chirped pulse analyzed in Section \ref{app2} could have indeed practical applications. As it was shown, the extended definition of the principal frequency depends on the chirp parameter $\beta$. As a consequence, we can envision the possibility to experimentally characterize Fourier-limited pulses in the few-cycle, single-cycle and sub-cycle regime. In fact, when the value of the chirp parameter approaches to zero the principal frequency reaches a maximum. Therefore, by looking for the central position of the third or fifth harmonics it should be  possible to determine if the pulse has a spectral phase (chirp), i.e., the blue-shift of the harmonics is maximum when the pulse is Fourier-limited.

Another attractive arena where the extended definition of the principal frequency might be relevant is the interaction of strong laser pulses with solid materials. Here, it was demonstrated that coherent radiation in the extreme XUV range can be generated driven bulk samples with sub-cycle pulses~\cite{luu2015extreme}, [Optica]. A correct characterization of this generated radiation could be indeed performed using the extended definition of the principal frequency. Another area of research would be to use SO pulses to drive solid samples. For this case, for instance, we expect the role of the intra and interband dynamics to be different compared with the case of `conventional' laser pulses. The extended definition of the principal frequency could shed light about the underlying physics of these strong field processes.

\section*{Acknowledgements}

M. F. C.~acknowledges financial support from the Guangdong Province Science and Technology Major Project (Future functional materials under extreme conditions - 2021B0301030005). E. G. N.~acknowledges to Consejo Nacional de Investigaciones Cient\'ificas y T\'ecnicas (CONICET). F. A. V.~ acknowledges to Comisi\'on de Investigaciones Cient\'ificas de la Pcia. de Buenos Aires.   

%$$\input{appendix.tex}\
\appendix*
\section{Principal Frequency and degree of fitting with SO functions}
%Zero crossing and peak values of SO functions
\label{app}

%Comentario (Lore): aca hay que referirse a E_1 y no a E en general, porque es ese campo y no otro el que se analiza, y asi lo llamamos en el cuerpo del trabajo. En la figura se puso E solo en el eje porque en la leyenda luego se aclara que campo particular es cada curva
We present here a numerical analysis of the SO frequency for the field $E_1(t)$ (Eq.~\eqref{e3}), at a fix value of the SO parameter $a$ (also called ``degree of super-oscillation" \cite{Berry_2006}), while varying the value of $N$.  
For each pair of values $\{N,a\}$, we compare the degree of fitting in the description of the peaks values (maxima and minima) and zero crossing points of the field $E_1(t)$, when the principal frequency, $\omega_{P1_\Phi}$, or the conventional SO frequency, $\omega_{SO}=a\omega_f$, are used. For that purpose, the real part of the field $E_1(t)$, with a global phase $\phi=\pi/2$, is compared against sinusoidal signals with frequencies $\omega_{P1_\Phi}$ and $\omega_{SO}$.

\begin{figure*}[!ht]
\centering
\includegraphics[width=0.8\textwidth]{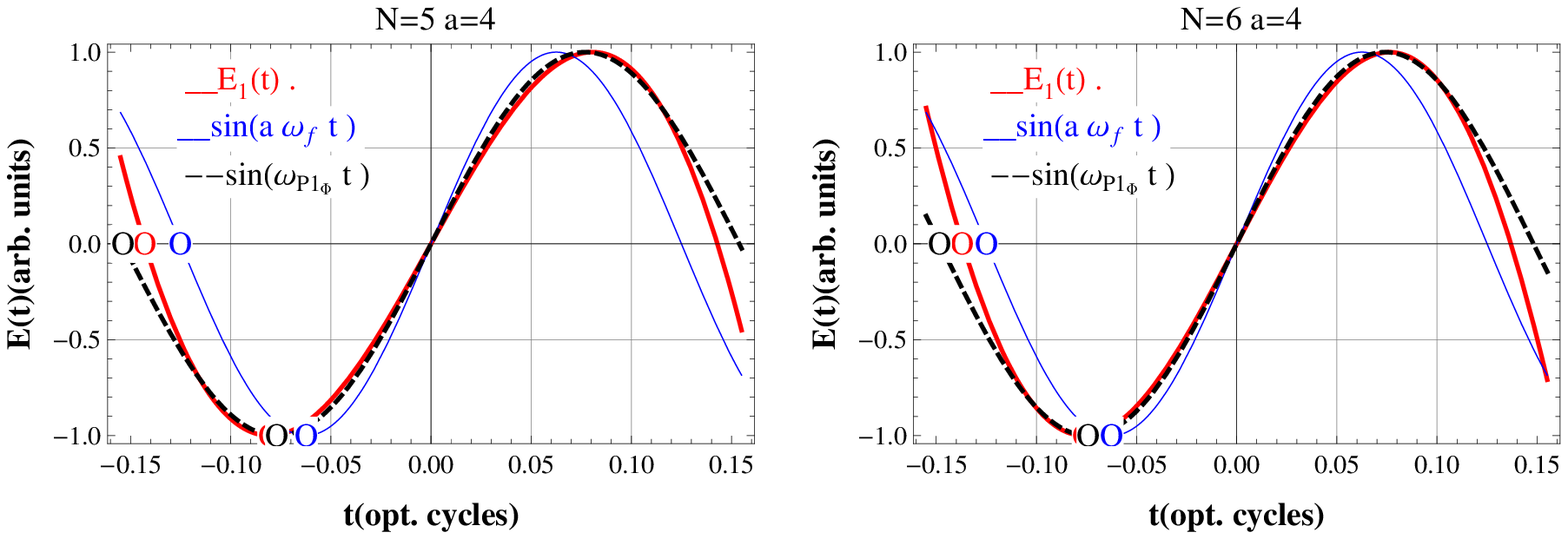}\\
\vspace{-2cm}
\includegraphics[width=0.8\textwidth]{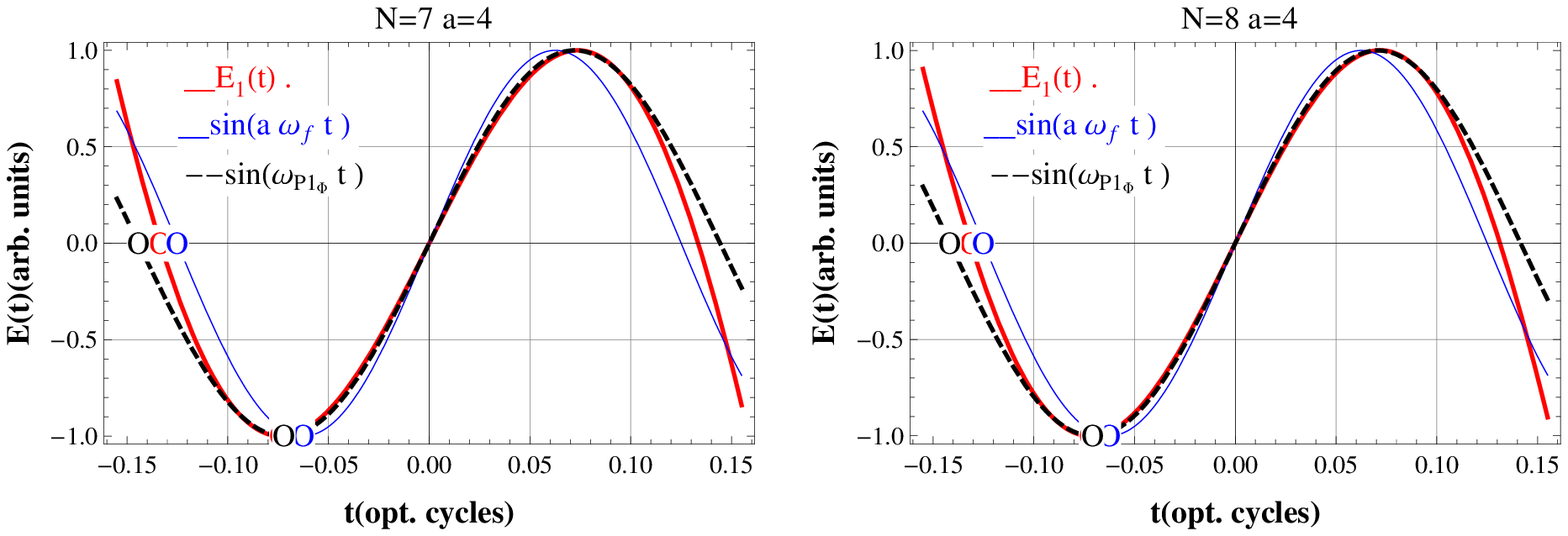}\\
\vspace{-2cm}
\includegraphics[width=0.8\textwidth]{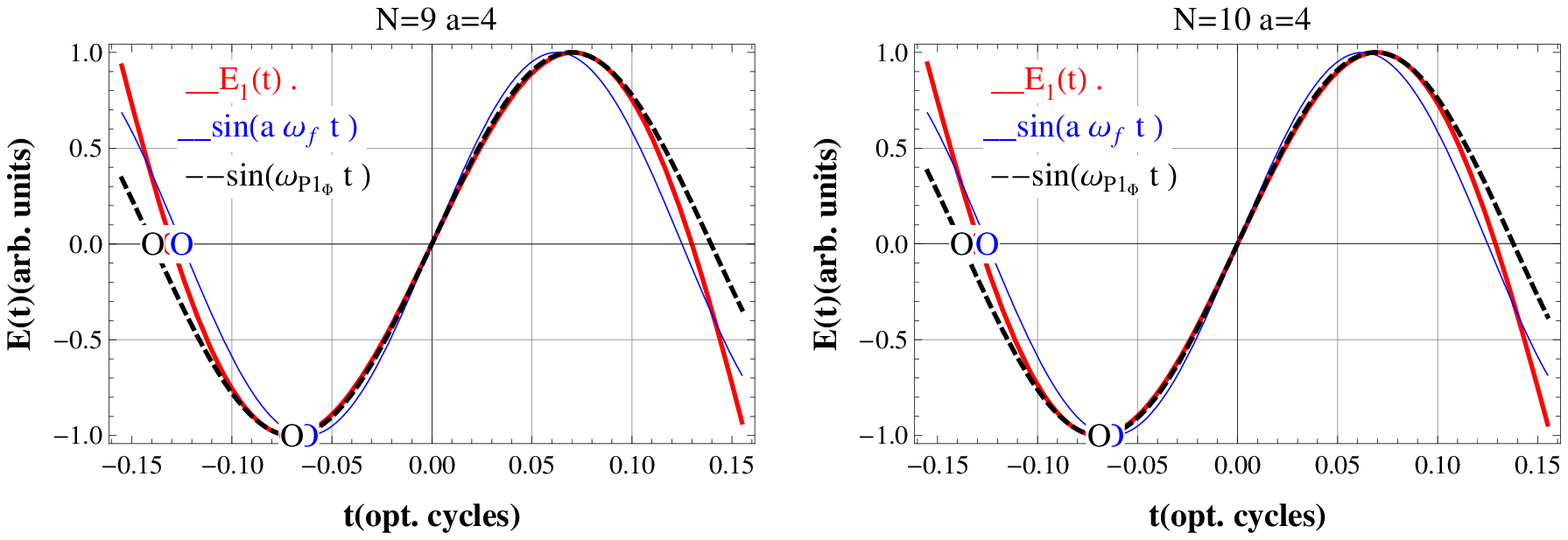}
\vspace{-1cm}
\caption{Comparison between zero crossing points and minima of the real part of the SO function $E_1(t)$ (red solid lines) and sine waves at the frequencies $\omega_{P1_\Phi}$ (dashed black lines) and %sine of the super oscillatory frequency 
$\omega_{SO}=a\omega_f$
(solid blue lines). The super-oscillatory degree, accounted by the parameter $a$, is hold at 4 for all curves, while $N$ varies from 5 to 10. It can be seen that the difference between $\mathrm{Re}[E_1(t)]$ and $\mathrm{Re}[e^{ia\omega_ft}e^{i\frac{\pi}{2}}]=\mathrm{sin}(a\omega_ft)$ is larger than that with respect to $\mathrm{Re}[e^{i\omega_{P1_\Phi}t}e^{i\frac{\pi}{2}}]=\mathrm{sin}(\omega_{P1_\Phi}t)$ . } \label{fig:app-harmonics1}
\end{figure*}

Figure~\ref{fig:app-harmonics1} depicts the curves of interest, $\mathrm{Re}[E_1(t)]$, $\mathrm{Re}[e^{i\omega_{P1_\Phi}t}e^{i\phi}]$ and $\mathrm{Re}[e^{ia\omega_ft}e^{i\phi}]$, when $a=4$, and $N$ ranges from $N=5$ to $N=10$. The time axis is displayed in units of optical cycles, normalized to $2\pi/\omega_f$, in a range that allows 
summarizing the main features of the fields in the SO region (it is the range where the first extreme values of the curves are localised).
The analysis is done on the positions of zeros and minima of the different curves (color circles), but the same conclusions could be obtained by analyzing the maxima, since maxima and minima, as well as the zero crossings, are localised symmetrically to $t=0$ (the minima in the interval $[-0.10, -0.05 ]$, the maxima in the interval $[0.05,0.10]$).
From these plots it can be observed that the agreement between $\mathrm{Re}[E_1(t)]$ (solid red lines) and $\mathrm{Re}[e^{i\omega_{P1_\Phi}t}e^{i\phi}]$ (black dashed lines) is better than that with $\mathrm{Re}[e^{ia\omega_ft}e^{i\phi}]$ (solid blue lines), considering the interval between the minimum and its adjacent maximum. We note, however, the latter field represents better the zero crossing points of $\mathrm{Re}[E_1(t)]$. % as $N$ increases. 
Besides, the fitting using both $a\omega_f$ and $\omega_{P1_\Phi}$ improves as $N$ increases. %Considering we are using an odd function, the same behaviour is observed for the maxima. 
Although we show here the analysis for a single value of the parameter $a$, this behaviour is universal.

\begin{figure*}[!ht]
\centering
\includegraphics[width=0.6\textwidth]{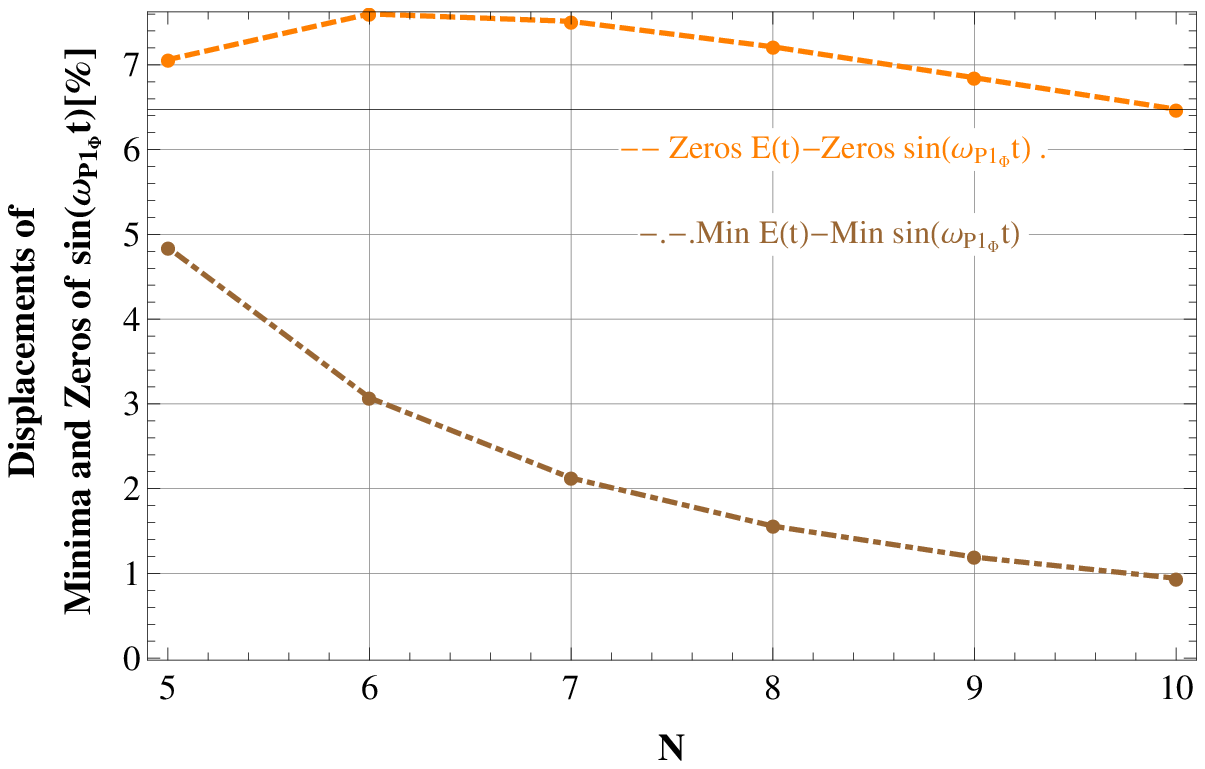}
\caption{Difference between the minima (brown dashed-dot lines) and zeros (orange dashed lines) of $\mathrm{Re}[E_1(t)]$ and those extracted from 
$\mathrm{Re}[e^{i\omega_{P1_\Phi}t}e^{i\frac{\pi}{2}}]=\mathrm{sin}(\omega_{P1_\Phi}t)$. These differences correspond to the curves shown in Fig.~\ref{fig:app-harmonics1}, where $a=4$ for all values of $N$. %The displacements between minima of $E_1(t)$ and $\mathrm{Re}[e^{i\omega_{P1_\Phi}t}]$  are smaller than displacements between zeros of the same functions, in agreement with the better fit around $t=0$.
}
\label{fig:app-harmonics2}
\end{figure*}

\begin{figure*}[!ht]
\centering
\includegraphics[width=0.6\textwidth]{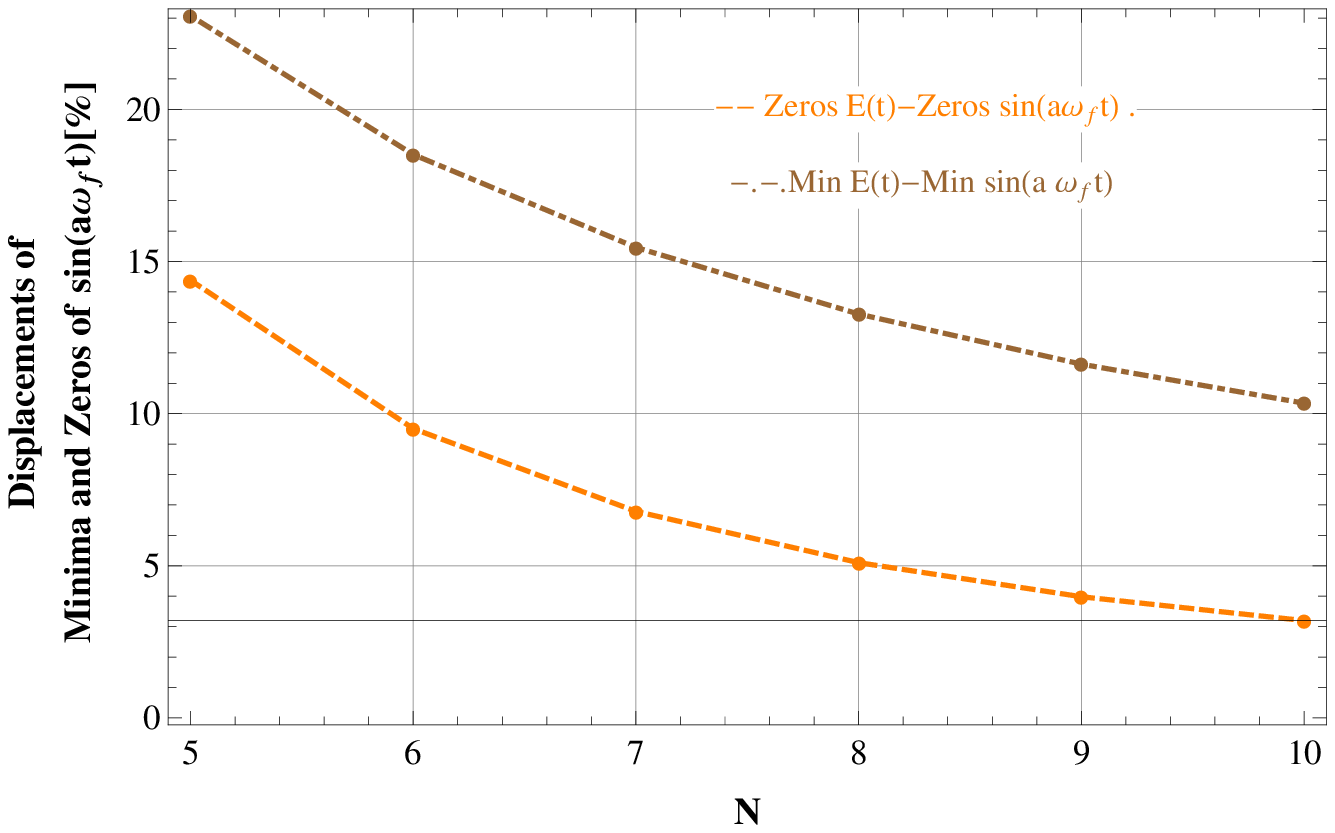}
\caption{
 Difference between the minima (brown dashed-dot lines) and zeros (orange dashed lines) of $\mathrm{Re}[E_1(t)]$ and those of 
$\mathrm{Re}[e^{ia\omega_ft}e^{i\frac{\pi}{2}}]=\mathrm{sin}(a\omega_ft)$. These differences correspond to the curves shown in Fig.~\ref{fig:app-harmonics1}, where $a=4$ for all values of $N$.}
\label{fig:app-harmonics3}
\end{figure*}

From the results of Fig.~\ref{fig:app-harmonics1}, we plot, in Fig.~\ref{fig:app-harmonics2}, the difference between the zeros (orange dashed lines) and minima (brown dot-dashed lines) of $E_1(t)$ and those of $\mathrm{Re}[e^{i\omega_{P1_\Phi}t}e^{i\phi}]$, when $N$ varies. 
Additionally, and for comparison purposes, we show in Fig.~\ref{fig:app-harmonics3} the curves corresponding to the difference between the zeros (orange dashed lines) and minima (brown dot-dashed lines) between $E_1(t)$ and $\mathrm{Re}[e^{ia\omega_ft}e^{i\phi}]$ as a function of $N$. %These critical points (minima and zeros), are highlighted with color circles. (Comentario (Lore): los minimos no se corresponden con puntos criticos de las curvas. Me parece que no es necesario ninguna aclaracion, la curva junto con lo que se dice despues es suficientemente clara) 
In the first case (Fig.~\ref{fig:app-harmonics2}), the difference between the zeros remains around $7\%$ (orange dashed line) for all $N$, while for the minima we observe a decrease from 5 to 1 $\%$ (brown dashed line) as $N$ increases. These differences,
for both the zeros and minima, are considerably smaller than those between $E_1(t)$ and a sine-wave at the SO frequency $a\omega_f$, $\mathrm{Re}[e^{ia\omega_ft}e^{i\phi}]$ (see Fig.~\ref{fig:app-harmonics3}). In the last case, the difference between zeros ranges from 15 $\%$ to 3 $\%$, while for the minima it is ranges from 23 $\%$ to 10 $\%$.

Finally, it must be mentioned that an analytical demonstration can be followed to show the convergence of $\omega_{SO}$ to $a\omega_f$, when $N\rightarrow \infty$~\cite{aharonov2017}. Moreover, we have shown, in Section~\ref{subsec:superoscillatory}, that $\omega_{P1_\Phi} = a\omega_f$ in such a limit ($N \rightarrow \infty$). Beyond of these analytical results, we show here that, for $N>1$ but finite, and for the first extreme values of the SO function $E_1(t)$, a sine-wave at $\omega_{P1_\Phi}$ is a best fit than a sine-wave at $a\omega_f$. This can be particularly recognized in Fig.~\ref{fig:app-harmonics1}, especially for the smallest values of $N$.

\providecommand{\noopsort}[1]{}\providecommand{\singleletter}[1]{#1}%

\end{document}